\documentclass[12pt]{article}

\input epsf
\def\beq{\begin{eqnarray}}
\def\eeq{\end{eqnarray}}

\def\lsim{\mathrel{\rlap{\lower3pt\hbox{\hskip0pt$\sim$}}
    \raise1pt\hbox{$<$}}}         %less than or approx. symbol
\def\gsim{\mathrel{\rlap{\lower4pt\hbox{\hskip1pt$\sim$}}
    \raise1pt\hbox{$>$}}}         %greater than or approx. symbol

\begin{document}

%\begin{flushright}
%NYU-TH/02/02/10 \\
%TPI-MINN-02/4\\
%UMN-TH-2044/02 \\
%\end{flushright}

\vskip 1cm
\begin{center}
{\Large \bf  Large  Hierarchies from Attractor Vacua}
%\vskip 0.2cm}

\vskip 1cm {Gia Dvali\footnote{\it  email:  dvali@physics.nyu.edu}}

\vskip 1cm
{\it Center for Cosmology and Particle Physics, Department of Physics, New York University, New York, NY 10003}\\
\end{center}

\vspace{0.9cm}
\begin{center}
{\bf Abstract}
\end{center}

We discuss a mechanism through which the multi-vacua theories, such as String Theory, 
could solve the Hierarchy Problem, without any UV-regulating physics at low energies. 
Because of symmetry the number density of vacua with a certain hierarchically-small
Higgs  mass diverges, and is an attractor on the vacuum landscape.  
 The hierarchy problem is solved in two steps. It is first promoted into a problem of the 
super-selection rule among the infinite number of vacua (analogous to $\theta$-vacua in QCD), 
 that are finely scanned by the Higgs mass.  This rule is lifted by heavy branes,
 which effectively convert  the Higgs mass into a dynamical variable.
 The key point is that a discrete "brane-charge-conjugation" symmetry
guarantees that the fineness of the vacuum-scanning is set by the Higgs mass itself.  
On a resulting landscape in all,  but a measure-zero set of  vacua the Higgs mass has a common hierarchically-small value. In minimal models this value is controlled by the  QCD  scale and is of the right magnitude.  Although in each particular vacuum there is no visible UV-regulating low energy physics, the realistic models are predictive.  For example, we show that in the minimal case the "charge conjugation" symmetry is automatically a family symmetry, and imposes severe restrictions 
on quark Yukawa matrices.

\vspace{0.1in}

\newpage

\section{Introduction}

It is assumed  usually that the solution to the Hierarchy Problem requires the existence of some 
new physics around  the TeV-scale, in order to regulate quadratic divergences in the Higgs
mass.  In the present article, we shall attempt to provide a counterexample to this statement.
Our approach will be based on generalization of \cite{dv}, which we shall closely follow. 
Another inspiration is provided by the recent progress \cite{gkp, ad, md, shamit} in understanding the distribution 
of vacua on String Theory moduli spaces. We will comment on the connection with the latter
work below.  

 In \cite{dv}  it was shown that in simple "string inspired" extensions of the standard model, postulating the coupling  to the heavy branes  and the three-form fields, the number of vacua with hierarchically small value of the Higgs VEV {\it diverges} due to symmetry reasons.  Such a value of the Higgs VEV was called  an "attractor" point in the space of vacua.  The crucial point is that by symmetry the attractor point is stable against the quantum corrections.  The situation is rather peculiar. 
The theory in question has an infinite number of vacuum states separated by the  large potential 
barriers. The Higgs mass (and the VEV)  takes different values in different vacua, but the vacua with the large
values of the Higgs VEV are rare and gradually increase in number density towards the smaller  Higgs VEV.  An infinite number density of vacua cluster around a certain hierarchically small value 
of the Higgs  VEV, which marks the attractor point.  In the other words, the scanning of the vacuum landscape by the Higgs mass becomes {\it hyper-fine} near the attractor value.  

   In each particular vacuum within the neighborhood of the attractor point, there is  no UV-regulating new physics around the scale of the Higgs mass. So a naive observer, living inside of any such vacua and suspecting nothing about the multiplicity of the similar vacuum states,  would attribute the smallness of the weak interaction scale to a mysterious fine tuning. 
Nevertheless, for an "outside" observer,  knowing that the number density of such vacua is divergent, the "fine-tunning" becomes perfectly natural.  The Hierarchy Problem is solved since the prior probability  distribution  is singular, and probability is sharply peaked around the vacua with the hierarchically small Higgs mass.

 These ideas indicate that multi-vacua theories, such as String Theory, have a potential of solving the Hierarchy Problem without invoking any UV-stabilizing  new physics around the weak scale. 
Hence, in such theories our ideas about naturalness must be reconsidered.  This view is supported
by the recent progress in the field. 

 It is becoming evident \cite{bp, gkp, kklt, lennie, bdg, ad, md, shamit, eva}  that the String Theory "landscape"\cite{lennie}, which reveals enormous complexity of the vacuum states, could play an important role in selection of 
the observed vacuum.  So it is conceivable that the attractor solution of the hierarchy problem could find a natural implementation in String Theory. 
Putting aside the question of the Higgs mass and the strength of the divergence in the vacuum number density,  the attractor of \cite{dv} can be viewed as some sort of a "holographic dual" to one class of vacua considered by 
Giddings, Kachru and and Polchinski \cite{gkp}, which arise at conifold points in Calabi-Yau moduli space\footnote{We thank Shamit Kachru for pointing out the possible connection, and for enlightening discussions. We also thank Michael Douglas for comments about the possible connection with his 
work \cite{md}.}. 
The recent interesting studies\cite{ad, md, shamit} indicate 
that "attractive" conifolds with flux can lead to clustering of high number of vacua. 
Although the clustering of the vacua in case of conifolds is weaker than in case of \cite{dv}, it will be shown below that this feature is rather parameter-dependent. 
This connection deserves further study \cite{shamit1}.  Some related ideas can also be found in \cite{eva}.

 Although, in the present paper we shall limit ourselves with effective field theory analysis of the attractor idea,  string theory concepts will play the crucial role in this analysis. 
We discuss how the attractor behavior arises in effective field theory that involves some of the key ingredients of string
compactifications,  such as the three-form fields, axions (two-forms) and effective 2-branes.  We shall show that, although the
attractor solution does not require any UV-regulating new physics, nevertheless the 
realistic models have a predictive power and are potentially testable.
For example, in the most minimal case, without any low energy extension of the Standard Model,  we show that the attractor-stabilizing symmetry is automatically  a {\it family} symmetry of Standard Model quarks, and implies restrictions on the structure of quark Yukawa couplings.   

 Among the possible approaches to the Hierarchy Problem, the  attractor solution is probably the closest possible analog of the axion solution of the Strong CP problem.
Indeed, from the beginning of our treatment the Hierarchy Problem, from being a problem of the
 UV-{\it stability} of the Higgs mass  ($m_{\phi}$), gets promoted into the problem of a {\it super-selection rule} among the infinite vacua scanned by $m_{\phi}$.  These are analogous to $\theta$-vacua in QCD that are scanned by the $\theta$ parameter\cite{theta}.
The advantage of dealing with a super-selection problem, as opposed to the one of UV-stability
is the following.  Unlike the latter problem,  the new physics that solves the former can be arbitrarily weakly coupled to
the Standard Model sector. 
The famous example of such new physics in case of the Peccei-Quinn (PQ) solution\cite{pq} of the Strong CP problem is the  axion
\cite{axion, axion1},    that  can be arbitrarily weakly coupled 
and practically invisible \cite{invisible, invisible1}.  As we shall see, in our case too, the new physics that solves the Hierarchy
Problem via attractor mechanism can be practically decoupled from the Standard Model sector. 
Of course, along with the analogies, there are fundamental differences with the 
axion mechanism.  In both cases the parameter of interest is promoted into a 
dynamical variable. In PQ case this is the $\theta$-angle, and in our case the Higgs mass $m_{\phi}$. 
In both cases, the vacua with small values of the parameters are selected, but the selection
mechanisms are different. In case of the $\theta$-angle the selection is via vacuum relaxation mechanism, since
small $\theta$ corresponds to the true groundstate of the system\cite{vw}. In our case, the selection
happens through the enormous multiplicity of states.  Because of  attractor,  in all, but a measure-zero number of vacua,  the Higgs mass is small.

The paper is organized as follows. In section 2, we briefly summarize our criteria of naturalness
and the essence of the attractor vacuum.  In section 3, we gradually build the attractor vacuum, 
by putting all essential ingredients together. At the end of this section we derive an effective equation
for finding the vacua on the landscape, and give their number counting.  
In section 4, we give a general discussion of the shift of the attractor point to the realistic values
of the Higgs VEV. The precise mechanisms are discussed in sections 5 and 6.  
In section 5, we study the effect of world volume terms, such as brane-localized mass terms, 
 on the attractor dynamics, and show that depending on the parameters they  can either 
lead to sharpening of the attractor, or to regulating it (that is, cutting-off the divergence in vacuum 
number densities).  In section 6,  we discuss the realistic model building  and present the two
versions of the complete models.  We show that  the minimal one, that  requires no enlargement of the electro-weak sector imposes severe  restrictions, on  the structure of Yukawa matrices. 
The next to minimal case, which is less constraining is discussed at the end of section 6. 
In section 7, we show that when extended to grand unified theories, the attractor solution 
automatically solves the problem of doublet-triplet splitting.  In appendix A, we show how the
theories with attractor can be obtained  from local
gauge-invariant  theories after integrating out the St\"uckelberg  field, and in appendix B we discuss
some related  potential issues.  In appendix C, we discuss some exact solutions.
In appendix D, we show, following  \cite{dv}, that the branes charged under the three-form fields
can be the axionic domain walls.  Finally in appendix E, we study how the axionic walls with field-dependent charges could  be obtained without the St\"uckelberg method, by integrating out  some intermediate scalars.

\section{Naturalness and the Essence of the Attractor Phenomenon}

In this section, we wish to briefly formulate our criteria of naturalness, and summarize the essence of the attractor 
solution.  The Hierarchy Problem is the problem of UV-sensitivity of the Higgs mass ($m_{\phi}$)
and consequently of the Higgs VEV ($\langle \phi\rangle$).  The excellent formulation of the problem can be found in \cite{lennie1}. 
The attractor mechanism solves this problem by selecting the vacuum with the hierarchically small Higgs mass. The selected value is UV-{\it insensitive}  due to the symmetry reasons.  In this respect, the attractor solution of the Hierarchy Problem is natural in the same sense as the 
Technicolor\cite{lennie1, savas}, or the low energy supersymmetry, but has some advantages over the latter. For example, as we shall show, the attractor mechanism also automatically solves the 
problem of Doublet-Triplet Splitting in grand unified theories,  which supersymmetry alone fails to solve. 

By simple choice of symmetry, the attractor solution gives possibility to obtain the Higgs VEV in terms of the QCD scale. 

 The essence of the attractor solution can be summarized as  follows. 
Attractor is a special  point on the energy landscape that contains an infinite number of discrete vacua. The vacua are scanned by a given parameter, which in the case of interest is the Higgss mass. 
What makes the attractor landscape special is the fact that, due to the symmetry reason, 
all, but the measure-zero set of the vacuum states exhibit  practically-equal and  hierarchically-small values of the Higgs mass.  This value is the attractor point.  

 The  attractor vacuum is achieved in two steps.  First, by coupling the Higgs to 
the field strength of an antisymmetric three-form, we promote the Hierarchy Problem into 
the problem of the super-selection rule, analogous to the Strong CP problem in QCD.  The 
vacua are scanned by the Higgs mass $m_{\phi}$, which plays the role analogous to  $\theta$-angle that scans the QCD-vacua.  The super-selection rule is lifted by 2-branes that effectively {\it source}
the Higgs mass, and allow the quantum transition between the vacua with different $m_{\phi}$. 
Thus, the Higgs mass gets  promoted into a dynamical variable.  This shares some analogy with 
the axion solution of the Strong CP Problem, in which $\theta$-angle gets promoted
into a dynamical variable. However, the reason why the desired vacuum is selected in the attractor 
case is different from the axion scenario. In the latter case, because the dynamical $\theta$-angle
changes continuously, the selection  is energetic. The selected vacuum is the true ground-state of the 
theory, to which the system relaxes on a microscopic time scale.  In our case, the transitions 
between the vacua with different $m_{\phi}$ are discretized. The barrier between the vacua is
very large, but the fineness of $m_{\phi}$- scanning changes throughout  the ladscape, and
becomes super-fine around the attractor  point. 
Because, of the large potential barrier,  each vacuum is extremely long lived. In such a situation energetics plays no role 
in selecting the vacuum. Instead what is important is the density of the vacua with the given values of $m_{\phi}$. 
 We show that due to symmetry reason, which triggers a profound back reaction on the brane charge, 
essentially all  vacua, cluster around a certain hierarchically-small value
of $m_{\phi}$, which is the attractor point.  

\section{The Attractor Vacuum}

\subsection{The Super-Selection Rule}

 Expanding the analysis of  \cite{dv},  we start  with a detailed discussion of the "attractor" idea. 
String Theory contains various antisymmetric form fields, which after compactification to
four dimensions give rise to three-forms, two-forms (axions) and one-forms (vectors).
The crucial role in our solution of the hierarchy problem is played by the 
three-form field $C_{\alpha\beta\gamma}$.  For a free three-form field the lowest order parity-invariant action 
has the following form
\begin{equation}
\label{action}
  \int_{3+1} \, {1 \over 48} \, F_{\mu\alpha\beta\gamma}F^{\mu\alpha\beta\gamma} 
\end{equation}
where $F_{\mu\alpha\beta\gamma}\, = \, d_{[\mu}C_{\alpha\beta\gamma]}$ is the four-form
field strength.  This action is invariant under the gauge transformation
\begin{equation}
\label{gauge}
C_{\alpha\beta\gamma}  \rightarrow  C_{\alpha\beta\gamma} \, + \, d_{[\alpha}\Omega_{\beta\gamma]},
\end{equation}
where $\Omega$ is a two-form. 
Because of this gauge freedom  in four dimensions $C$ contains no propagating degrees of freedom, 
and its equation of motion
\begin{equation}
\label{fequation}
\partial^{\mu}F_{\mu\nu\alpha\beta} \, = \, 0 
\end{equation} 
is solved by 
\begin{equation}
\label{solution}
F_{\mu\nu\alpha\beta} \, = \, F_0 \,  \epsilon_{\mu\nu\alpha\beta}, 
\end{equation} 
where $F_0$ is an arbitrary constant. 
Hence, in the absence of other interactions, the effect of the 3-form is  reduced to adding
an arbitrary integration constant to the Lagrangian. This constant  will contribute to the overall cosmological term.    
However, in the presence of interactions with the other fields, the integration constant  $F_0$ will also contribute to their effective masses and couplings. Consider, for instance, an interaction with a scalar field 
$\phi$. In what follows we shall treat $\phi$ as the prototype of the Standard Model Higgs. 
The lowest order parity and gauge-invariant Lagrangian describing a non-trivial interaction 
between $\phi$ and $C$ has the following form
\begin{equation}
\label{fphi}
L \, = \, |\partial_{\mu}\phi|^2 \,  - \,  {1 \over 48} F^2 \, + \,  \, |\phi|^2 \, \left (m^2\,  + \,  {F^2 \over 48M^2}
\right) \, - \, {\lambda \over 2} \, |\phi|^4
\, + \, ...
\end{equation}
Here $\lambda$ is the quartic coupling, and $m$ and $M$ are the mass parameters, that naturally are of the order of the UV cut-off, which we shall take to be around the Planck mass $M_P$.  For definiteness,  
we shall assume $m^2 \, > 0,\,  M^2 > 0$.  Note that the value of $F$ determines the value of the effective mass$^2$ and consequently the VEV of the Higgs field. The latter is equal to
\begin{equation}
\label{vev}
|\phi|^2 \, =\, {1 \over \lambda} (m^2 \, + \, F^2/48M^2)   
\end{equation}
The value of $F$ is determined from the equation of motion
\begin{equation}
\label{fequation}
\partial^{\mu}\left ( (1 \, - \, |\phi|^2 /M^2 )\,  F_{\mu\nu\alpha\beta} \right ) \, = \, 0, 
\end{equation} 
which is solved by 
\begin{equation}
\label{slutionfhiggs }
F_{\mu\nu\alpha\beta} \, = \, {F_0 \,  \epsilon_{\mu\nu\alpha\beta} \over (1 \, - \, |\phi|^2 /M^2 )}
\end{equation}
where $F_0$ is an arbitrary integration constant.
Plugging this solution into the equation for the Higgs field we get  the following effective 
equation determining the Higgs VEV
\begin{equation}
\label{higgseq}
 \left (- m^2\,  + \,  {F_0^2 \over 2M^2 (1 \, - \, |\phi|^2/M^2)^2} \right) \phi \, +  \, {\lambda } \, |\phi|^2\phi\, = \, 0
\end{equation}
It is obvious that the above theory has a continuum of the vacuum states, labeled by $F_0$. 
In many of these vacua the VEV and the mass of the Higgs is much smaller than the 
cut-off. These are the vacua with 
$m^2 - F_0^2/2M^2\, \ll M^2$, in which $\phi \ll M$. 
For instance, there is a vacuum 
with $F_0^2\, = \, 2m^2M^2$ in which $\phi\, = \, 0$. In this vacuum $\phi$ is  exactly massless.

Although, there exist vacua with a
light scalar,  the hierarchy problem  is  nevertheless {\it not} solved in the above theory. 
The reason is twofold.  First in the above theory the vacua are uniformly scanned by the integration
constant  $F_0$, and the light Higgs vacua  are not special in any way. 
Secondly,  $F_0$ is not a dynamical quantity, and there is no transition between 
the different vacua.  In the other words there is a super-selection rule in $F_0$,  no  vacuum is preferred over any other, and any choice of $F_0$ is good. 
In this respect $F_0$-vacua are similar to theta-vacua in QCD\cite{theta}.
% where the vacua with small value of the $\theta$-angle are not particularly preferred by the system. 
As it is well known,  in QCD with no massless quarks there is a continuum of {\it physically distinct}
vacuum states, that can be parameterized by a periodic variable $\theta$.  These $\theta$-vacua obey the super-selection rule, there is no transition between the states with different $\theta$. This situation gives rise to celebrated  Strong CP Problem, since the phenomenologically acceptable
vacua,  with $\theta < 10^{-9}$ are not particularly preferred by the system. 

 The situation in our case is analogous. The model (\ref{fphi}) has a continuum of the vacuum states, scanned by $F_0$ or equivalently by  $\langle \phi \rangle$, and there is no transition between the 
different $\langle \phi\rangle$-vacua.  Thus, the hierarchy problem, from the problem of UV-instability 
of the Higgs mass, got promoted into the super-selection problem, analogous to 
the Strong CP Problem in QCD. 
 
 In order to solve the former, we shall try to follow, as closely as possible, the general strategy adopted 
by the axion solution of the Strong CP Problem. As it is well known, this solution is based on, {\it a)} first promoting $\theta$ into a dynamical variable (axion),  and {\it b)} showing that  $\theta=0$ is the true groundstate of the system.  

 Thus, in order to solve the hierarchy problem we need to accomplish the two more steps. 

$~~~$

{\bf {\it 1)}} Promote $F_0$ into a dynamical variable; 

$~~~$

{\bf  {\it 2)}}  Find the symmetry reason that will ensure that the vacua are not uniformly distributed and the
vacua with the light Higgs are preferred over all the others.  

$~~~~$

As we shall see, the first step is achieved by introduction of branes which permit the transition between the different $\phi$-vacua.  The difference with the axion case is that the transition is quantum-mechanical and discrete, as opposed to being classical and continuous.  This circumstance creates a profound difference between our 
solution of the Hierarchy Problem  and the axion solution of the Strong CP one.  
Impossibility of the continuous classical transition tells us that we cannot directly generalize the 
axion mechanism of $\theta$-relaxation, but it also suggests a natural substitution. Indeed, because there is
no classical transition, the energy difference is unimportant for selecting the vacuum state, and what matters is the multiplicity of vacua with a given value of $\phi$.  We shall see the hierarchy problem is solved, because infinite number of vacua cluster around the small $\phi$-value, by the symmetry reason.  

 In what follows we shall give a detailed discussion of this phenomenon.

\subsection{The Role of the Branes:  Breaking the  Super-Selection}

 As said above, the super-selection rule is lifted  by introducing 2-branes (membranes) that source our three-form field
$C_{\alpha\beta\gamma}$.  Ignoring the Higgs field for a moment, the effective action incorporating the
interaction with branes can be written as
\begin{equation}
\label{qbrane}
{q\over 6}\, \int_{2+1}  d^3\xi \, C_{\mu\nu\alpha} \left( {\partial Y^{\mu} \over \partial \xi^a}
{\partial Y^{\nu} \over \partial \xi^b}{\partial Y^{\alpha} \over \partial \xi^c}\,\right) \epsilon^{abc}
   \, - \,  \int_{3+1} \, {1 \over 48} F^2, 
\end{equation}
where the first explicitly-written term describes the interaction between the brane and the three-form.
$q$ is the charge of the brane, and $x^{\mu} \, = \, Y^{\mu}(\xi)$ specify a $2+1$-dimensional history
of the brane in $3+1$-dimensions as a function  of its world-volume coordinates $\xi^a ~(a=0,1,2)$. 
This term can be rewritten in form of the following four-dimensional integral 
\begin{equation}
\label{CJcoupling}
\int \, d^4x \,  {1 \over 6} J^{\alpha\beta\gamma} C_{\alpha\beta\gamma}
\end{equation}
where $J^{\alpha\beta\delta}$ is the brane current 
\begin{equation}
\label{current}
J^{\alpha\beta\gamma}(x)\,  = \,  \int d^3\xi \delta^4(x \, - \, Y(\xi))\, q \,  
 \left( {\partial Y^{\alpha} \over \partial \xi^a}
{\partial Y^{\beta} \over \partial \xi^b}{\partial Y^{\gamma} \over \partial \xi^c}\,\right) \epsilon^{abc} 
\end{equation} 
Obviously, the current $J_{\alpha\beta\gamma}$ is conserved as long as $q$ is a constant. 

 The brane self-action has the  standard form
\begin{equation}
\label{branetension}
 - T\int \, d^3\xi \sqrt{-g},
\end{equation}
where $T$ is the brane tension (a mass per unit surface), and  $g_{ab} = \partial_aY^{\mu}\partial_bY^{\nu}\eta_{\mu\nu}$ is the induced metric on the brane. 
Note that, since the bulk $4$-dimensional gravity plays no essential role in our considerations, we 
have taken a flat Minkowskian  $4$-dimensional metric $\eta_{\mu\nu}$.  Despite this, 
the induced metric on the brane is not flat in general, due to the dynamical curving of the brane.
 With the brane source taken into the account,  the equation of motion of the 3-form now becomes, 
\begin{equation}
\label{feqy}
\partial_{\mu}\,  F^{\mu\nu\alpha\beta} \, = \, -\, q\int d^3\xi \delta^4(x \, - \, Y(\xi)) 
 \left( {\partial Y^{\nu} \over \partial \xi^a}
{\partial Y^{\alpha} \over \partial \xi^b}{\partial Y^{\beta} \over \partial \xi^c}\,\right) \epsilon^{abc} 
\end{equation} 
The brane can be taken to be flat and static,  $Y^{\mu} \, = \, \xi^{\mu}$ for $\mu = 0,1,2$, and 
$Y^3 \, = \, 0$. 
The equation of motion then simplifies to  
\begin{equation}
\label{feqstatic}
\partial_{\mu}\,  F^{\mu\nu\alpha\beta} \, = \, -\, q \delta(z) \epsilon^{\nu\alpha\beta z} 
\end{equation} 
where $z=0$ is the location of the brane. 
Both (\ref{feqy}) and (\ref{feqstatic}) show that the brane separates the two vacua 
in either of which  $F_0$ is constant, and the two values differ by $|q|$.
 Thus, the introduction of branes ensures that the transition  between the vacua with different values of $F_0$  is possible, as long as the value of $F_0$ changes by the integer multiple of $q$.
Hence the discrete quantum transition between the different vacua are possible via nucleation of 
closed branes (this fact was used in an interesting attempt\cite{bt} to explain the smallness  of the cosmological term).  
 
In the other words, the theory given by the action (\ref{qbrane})  has multiplicity of vacua that can be labeled by an integer $n$. The value of the field strength in this vacua is  
\begin{equation}
\label{fzeron}
-{1\over 24}F_{\alpha\beta\gamma\mu}\epsilon^{\alpha\beta\gamma\mu} \, =\, F_0 \, = \, qn \, + \, f_0, 
\end{equation}
where $f_0$ is a constant, which we will set equal to zero.  
 That is, the value of $F$ is quantized in units of the brane 
charge . 
 Restoring the coupling to the Higgs field, the equation (\ref{higgseq}) determining the Higgs VEV
now becomes
 \begin{equation}
\label{higgseqn}
 \left (- m^2\,  + \,  {(nq)^2 \over 2M^2(1  \, - \, {|\phi|^2\over M^2} )^2 } \right) \phi \, +  \, {\lambda } \, |\phi|^2\phi\, = \, 0
\end{equation}
The good news is that now the transition between the vacua with different $n$ are possible, however
the hierarchy problem is still not solved.  First,  because both $m$ and $M$ are large, we need a very 
small $q$ in order to ensure a fine enough scanning of the Higgs mass. 
Secondly, even for a small $q$, the vacua with a small Higgs VEV are not preferred over the others. 
Both problems can be cured in one shot, by requiring a symmetry which will promote 
$q$ into a continuous function of $\phi$.  For example,
\begin{equation}
\label{qattractor}
q\, \rightarrow \, q(\phi) \,  \propto \, \phi^N
\end{equation}
 Then, the zero of the function $q(\phi)$ will become  an "attractor" (accumulation of infinite number
of vacua) in the space of vacua. 
This will guarantee both the super-fine scanning 
of the Higgs mass and the preference of the vacua with the small values of the Higgs VEV.
Thus,  idea of the "attractor" is that the multiplicity of the vacua  with the small values of the Higgs VEV
become divergent (or at least very sharply peaked), because of the symmetry reasons.

 The essence of the attractor phenomenon can be summarized schematically in the following sequence 
(written in units of the fundamental scale)
\begin{equation}
\label{sequence}
(q=\phi^N) \, \rightarrow \, (\Delta F = q) \, \rightarrow \, \left({\Delta \phi \over \phi}  \sim q\right )  \, \rightarrow \,
(\Delta q \sim q \phi^{N-2})
\end{equation}
The arrows indicate that a non-zero charge of the brane separating the two different  vacua  implies the change of $F$ across the brane, which implies the change of $\phi$. The latter implies the change of the brane charge  in a new vacuum, and closes the cycle.  
We shall give a detailed discussion of the dynamics of the above  sequence throughout the paper. 

\subsection{Charge Conjugation: Creating an Attractor}

 Thus, to achieve an attractor,  we must promote the brane charge $q$ into the function of $\phi$, such that $q$ vanishes for 
small values of $\phi$.  To guarantee this, we shall require a new 
discrete symmetry $Z_{2N}$
\begin{equation}
\label{conjugation}
\phi \rightarrow e^{i{\pi\over N}}\phi,
\end{equation}
which acts on the brane as "charge conjugation". That is, we require that 
$q\rightarrow -q$ under $Z_{2N}$.  The invariance under $Z_{2N}$ then demands 
$q$ to be an odd function of $\phi^N$. The simplest choice is 
\begin{equation}
\label{charge}
q \, = \, q_{eff}(\phi)\, = \, {\mu \over 2} \left ({\phi \over M_P} \right )^N\, + \, h.c..   
\end{equation}
where $\mu$ is some constant.

 For successful implementation of the above idea, we need to address the following technical issue. 
For non-constant $q$ the current (\ref{current}) is not conserved, and hence the coupling 
(\ref{CJcoupling}) is not gauge invariant.   In order to maintain the gauge invariance, 
following \cite{dv}, we shall modify this coupling in the following way.
\begin{equation}
\label{newcoupling}
-\,\int \, d^4x \,  {1\over 6} \,C_{\alpha\beta\gamma} J_{(T)}^{\alpha\beta\gamma}
\end{equation}
where $J_{(T)}$ is the transverse part of the current
\begin{equation}
\label{transverse}
 J_{(T)}^{\alpha\beta\gamma}\, = \, {1\over 6} \, \Pi^{[\alpha}_{\mu}J^{\mu\beta\gamma]}.
\end{equation}
Here 
$\Pi_{\mu\nu}\, = \eta_{\mu\nu} \, -\, {\partial_{\mu}\partial_{\nu} \over \partial^2}$ is the transverse projector. \footnote{$J_{(T)}$ can also be written in the following form 
$J^{\alpha\beta\gamma}_{(T)} \, = \, {1 \over 6} \, \epsilon^{\alpha\beta\gamma\mu}\Theta^{\nu}_{\mu} \epsilon_{\nu\tau\rho\omega} C^{\tau\rho\omega}$, where $\Theta_{\mu\nu}\, = \, {\partial_{\mu}\partial_{\nu} \over \partial^2}$ is the longitudinal projector.}

 For constant $q$, we have  $\partial^{\alpha} J_{\alpha\beta\gamma}\, =\, 0$ and $J_{(T)} \, = \, J$.
Thus, the coupling (\ref{newcoupling}) reduces to (\ref{CJcoupling}).
 This fact accomplishes our goal.
In each given vacuum the expectation value of Higgs is fixed, and so $q$ is constant.  So in each vacuum  with unexcited Higgs field the brane  couples to the three-form in a normal way. 
On the other hand, the change of $q$ from vacuum to vacuum is permitted, because $C_{\alpha\beta\gamma}$ only couples to the transverse part of $J_{\alpha\beta\gamma}$. 
The existence of the attractor point at $\phi=0$ is guaranteed by the fact that  $J_{(T)} \rightarrow 0$  when $\phi \rightarrow 0$. 

The coupling (\ref{newcoupling}) is the gauge-invariant generalization of (\ref{CJcoupling}) for the
case of a non-constant charge $q(\phi)$. 
For the constant $\phi$ the above coupling is equivalent  to (\ref{qbrane}) with $q_{eff} \, = \, \mu \, Re(\phi/M_P)^N$.  Although, the coupling (\ref{newcoupling}) contains a projector, 
it actually can be obtained from a  local underlying theory after integrating out certain degrees of freedom. This issue is  discussed in the Appendix A.  It is shown there how the coupling (\ref{newcoupling}) can be obtained by integrating out the Goldstone-type degrees of freedom ala St\"uckelberg.  In a very crude sense, the St\"uckelberg field plays  the role analogous to the one of 
axion in the solution of the Strong CP problem. 
Some alternatives to St\"uckelberg method will be discussed in \cite{giga} \footnote{We thank Gregory Gabadadze for enlightening discussions on this and other issues.}.

 Putting all the ingredients together let us now show that with $Z_{2N}$-symmetry, theory has an attractor point in the space of vacua at $\phi=0$.  For the convenience we write down the combined Lagrangian, which takes the 
following form 
\begin{eqnarray}
\label{theaction}
L \, &=& \, |\partial_{\mu}\phi|^2 \,  - \,  {1 \over 48} F^2 \, + \,  \, |\phi|^2 \, \left (m^2\,  + \,  {F^2 \over 48M^2}
\right) \, - \, {\lambda \over 2} \, |\phi|^4 \\ \nonumber
&&\,- {1\over 6} \,C_{\alpha\beta\gamma} J_{(T)}^{\alpha\beta\gamma}
\end{eqnarray}
The above theory admits the  divergent number of vacua at small VEV of
$\phi$, at least when\footnote{The meaning of this constraint is to avoid "overshooting" to the 
vacua with the large positive Higgs mass,  in case if we start in a vacuum with $F=0$, in which the Higgs VEV is maximal, and so is the brane charge.} 
\begin{equation}
\label{cond}
m \gg {\mu \over \sqrt{2}M} ({m \over \sqrt{\lambda} M_P})^N.
\end{equation} 
In order to see this, we shall integrate out the brane and the 3-form field and write down the
effective potential for $\phi$.
Choosing the brane to be located at $z=0$, the equations are
\begin{equation}
\label{firstgrad}
\partial^{\mu} \left ( ( 1 -  |\phi|^2/M^2) F_{\mu\alpha\beta\gamma} \right) \, =
 \, \mu\epsilon_{\alpha\beta\gamma\nu} \Theta^{\nu z} \left [Re \left ({\phi\over M_P}\right ) ^N \, \delta(z)
\right ] 
\end{equation}
\begin{equation}
\label{secondgrad}
\partial^2 \phi \, - \, \left (\, m^2\, + \, {F^2 \over 48M^2}\right ) \, \phi \, + \, \lambda\, |\phi|^2\phi\, 
+ \,  {\mu N \over 12} {\phi^{*N-1} \over M_P^N}\, \delta(z) \, \Theta_{z\nu} \epsilon^{\nu\alpha\beta\gamma}\, C_{\alpha\beta\gamma}\, = \, 0
\end{equation}

In integrating these equations, for simplicity,  we will first make the following approximation which is well justified in the vacua with the small VEVs of $\phi$ (more rigorous derivation is discussed in the Appendix C).  

$~~~~~~~$

 {\it 1)} We shall ignore terms of order $\phi^2/M^2$ 
in the gauge-kinetic function of $F$ in l.h.s. of  equation (\ref{firstgrad}); 

$~~~~$

{\it 2) } We shall ignore the terms proportional 
to the derivatives of $q_{eff}$ on r.h.s. of (\ref{firstgrad}).

$~~~~$

 The above is justified, because the change in $q$ in each elementary transition between the small $\phi$ vacua is very small. So the correction to $\Delta F_0$ because of change in $q$ is 
of the higher order smallness in $\phi/M_P$, and can be ignored. 

Indeed,  change in $\phi$
($\Delta\phi$), in each elementary step that connects the neighboring vacua, is due to change of 
$F$, which is $\Delta F \simeq q_{eff}$.  Because the VEV of $\phi$ in any vacuum is 
given by (\ref{vev}), the change in $\phi$ is 
\begin{equation}
\label{deltaphi}
\Delta (\phi^2) \, \simeq  {\Delta F_0 F_0 \over \lambda M^2} \, \simeq \, q_{eff} {F_0 \over \lambda M^2}
\end{equation}
Thus, even if the attractor point happens to be for the maximal value $F \sim M^2$, the change
$\Delta \phi^2 \sim q_{eff}$, and consequently
\begin{equation}
\label{deltaq}
\Delta q_{eff} \, = \, {N\mu\over 2} {\phi^{N-2} \over M_P^N}\Delta(\phi^2) \, \sim \, q_{eff}\, {\mu \phi^{N-2} \over M_P^N}
\end{equation}
Thus,  the change in the brane charge is higher order in $\phi/M_P$.  So around the attractor point
($\phi \rightarrow 0$) this subleading correction can be safely ignored in each elementary step
and $q_{eff}$ can be regarded as constant.  Only after many steps the accumulated change in 
$\phi$ can become significant. 

This fact simplifies the equation (\ref{firstgrad}). Integrating it for the constant $q_{eff}$ we get that $F$
is given by (\ref{solution}) with 
\begin{equation}
\label{fn}
F_0 \, = \, nq_{eff}, 
\end{equation}
where $n$ is an integer that labels the different vacua. 
Plugging this result into the equation for $\phi$ we get the effective equation defining the 
VEV of $\phi$ \footnote{Without the loss of generality we can put VEV of $\phi$ to be real}  
\begin{equation}
\label{pot}
  - \left (m^2 - n^2{\mu^2 \over 2 M^2} \,\left (\phi^N \over M_P^N\right )^2\right ) \phi \, + \, \lambda \, \phi^3\,  = \, 0
\end{equation}
Thus,  $n$ labels different vacua.  It is obvious that there are infinite number of vacua 
close to $\phi = 0$.  Thus,  vacuum with a vanishing VEV is an {\it attractor}. 

 We should stress that $Z_{2N}$-symmetry guarantees the UV-stability of the attractor point. 
Indeed, the attractor point is the vacuum in which the brane charge vanishes $q \, =\, 0$. 
Due to $Z_{2N}$-symmetry, this happens when $\phi=0$. Thus, any renormalization  of $\phi$ implies the corresponding renormalization of  the brane charge, such that $\phi=0$ remains an attractor point.  

\subsection{Counting the Number of Vacua}

We shall now give a simple general rule for counting the number of vacua near the attractor point. 
For simplicity we set the parameter $\lambda \sim 1$.   We wish to estimate the
number of vacua in which the expectation value of the Higgs field is of the order of a given value 
$\phi_0 \ll M$. The number of the vacua in which the Higgs VEV is 
$\phi \sim \phi_0$,  we shall denote by $n_{\phi_0}$.
There is a simple way to estimate this number. We start with a vacuum with a given VEV
$\phi = \phi_0$ and begin lowering $\phi$ by jumping to new vacua via creating branes, till
the VEV changes in first non-zero digit.  
The VEV of $\phi$ in each vacuum is given by (\ref{vev}).
The change of $\phi$ in each elementary step is given by (\ref{deltaphi}) and is small, because $q$
is small.  The number of vacua $n_{\phi_0}$ is equal to the number of steps that will make the accumulated relative change of $\phi$  of order one.  That is $n_{\phi_0}$ is defined by the condition that the quantity
\begin{equation}
\label{relchange}
{\Delta \phi \over \phi_0} \, \sim  \, {qn_{\phi_0} \over \phi_0^2} 
\end{equation}
should become of order one. 
Thus, the number of vacua is 
\begin{equation}
\label{nphi}
n_{ \phi_0} \, \sim \ {\phi_0^2 \over q} \, \sim \,  {M_P^N \over \mu\phi_0^{N-2}} ,
\end{equation}
which for $N>2$ diverges as $\phi_0 \rightarrow 0$.

The above expression should not create a false impression that there is no attractor for 
$N=2$ or less.  The equation (\ref{nphi}) indicates that the number of vacua in which  
$\phi$ is of the order of a given value $\phi_0$ diverges as $\phi_0 \rightarrow 0$, as long as 
$N>2$. For smaller $N$, this number is either constant or decreases with  $\phi_0$, but the 
total number of vacua accumulated around $\phi=0$ is always infinite!

In the other words, the number of vacua with the Higgs VEV in an interval 
$M_P > \phi > \phi_0$  is given by the sum
\begin{equation}
\label{sum}
n_{(>\phi_0)} \, =\, \sum n_{\phi_0}
\end{equation}
where the sum is taken over all the discrete vacua.  As said above, for the condition (\ref{cond})  the vacua can be arranged
into the groups labeled by $\phi_0$.  In any vacuum belonging to $\phi_0$-th 
 group the VEV of the Higgs is equal to $\phi_0$ in the first nonzero digit, and the number in each group is given by $n_{\phi_0}$.   Although for $N=2$ or less this number either stays constant or decreases with $\phi_0$, the density of the groups grows as $1/\phi_0^2$
and over-compensates the decrease of $n_{\phi_0}$ for any $N>0$.  Putting it shortly, for the small $\phi_0$  the sum can be approximated by the following integral
\begin{equation}
\label{nvacua}
n_{(>\phi_0)} \sim  M_P\, \int_{\phi_0}^{M_P} \, n_{\phi} {d\phi \over \phi^2}
\end{equation}
which shows, that even for $N=1$ the total number of vacua diverges as log$(\phi_0)$.
The divergence of $n_{(>\phi_0)}$ for $\phi_0 \rightarrow  0$ for  $N>0$, can also be seen from (\ref{pot}),
which for  $N = 1$ gives the following expression  for the VEV of $\phi$ 
\begin{equation}
\label{N1}
|\phi_0|^2\, =\, {m^2 \over \lambda \, + \, n^2(\mu^2/2M^2M_P^2)} .
\end{equation}
This VEV approaches $\phi_0 = 0 $ for $n \rightarrow \infty$.

\section{The Weak Scale Shift of the Attractor}

We have shown that a theory in which the brane charge $q$ is set by the VEV of a scalar field
has a divergent number of vacua with the values of the scalar field for which $q \, \rightarrow \, 0$.
This is the  essence of the attractor phenomenon.
In the above example the "charge conjugation" $Z_{2N}$ symmetry guaranteed  that the 
attractor point was at $\phi\, = \, 0$. 
 However, in order to solve the hierarchy problem in the Standard Model we have to make sure that 
the attractor point  is not at zero, but instead at the observed value $\phi \sim$ 100GeV.  As it was shown in \cite{dv}, when the attractor theory is considered in the cosmological context of an eternally  inflating  Universe, the attractor point can be shifted to the value of the
Hubble parameter during inflation.
The solution of the hierarchy problem then would require  this parameter to be around the weak scale or so.  

  Despite the cosmological possibilities, it is important to have other mechanisms that  could generate the small shift of the attractor point in a cosmology-independent way.  One universal possibility is to shift the brane charge by a small $\phi$-independent amount 
\begin{equation}
\label{qnew}
q_{eff} \rightarrow  {\mu \over 2} \left ({\phi^N \over M_P^N }  \, - \, \xi\right )\, + \, h.c..   
\end{equation}
where $\xi$ is a $\phi$-independent part.  We should think of $q_0$ as of  "spurion", the VEV 
that  spontaneously (softly) breaks $Z_{2N}$ symmetry.  It is important to stress  that even 
for constant $\xi$,  because $\xi = 0$ is an enhanced symmetry point,  the value  of $\xi$ is  {\it perturbatively-stable}.  The attractor point  then will be shifted to 
\begin{equation}
\label{pnew}
\phi  \, = \, (\xi)^{{1\over N}}M_P
\end{equation}
The origin of $\xi$ is model dependent.  We shall explore two possibilities. 
In section 5, we will show that $\xi$ can come from the QCD condensate of the Standard Model quarks.\footnote{ $\xi$ can also be set by the expectation value of some hidden sector
fermionic condensate, which has no direct interactions to Standard Model particles.} 
 Another avenue, to be discussed in the next section, is to take into account the possible impact
of  
brane-localized mass term of the Higgs field.  As we will show, such terms can have non-trivial effects
on the attractor dynamics.

\section{The Effect of the Brane-Localized  Potential}

$\phi$ field may have various potential terms on the brane world volume, compatible 
with symmetries. The most important of these is a brane-localized mass term, which can be
introduced in the four-dimensional action in the following form
\begin{equation}
\label{massp}
-\,\int \, dx^4 \,  M_{br}(x)^2|\phi|^2,
\end{equation}
where 
\begin{equation}
\label{branemassterm}
M_{br}^2(x)\, =\, \pm \int\, d\xi^3\sqrt{-g}\, M_B \delta^4(x-Y).
\end{equation}
In the above expression $M_B$ is a positive mass parameter. 
The question is what is the effect  of this brane-localized mass term on the attractor dynamics? 
 We shall show that this effect depends on the sign in (\ref{branemassterm}). 
For the positive sing the attractor at $\phi=0$ becomes sharper,  meaning that the number of vacua diverges faster for  $\phi\rightarrow 0$.  In the case of the negative sing, the two sub-regimes are possible, depending on the parameters.  One possibility is that the attractor point  is shifted away from zero, but the divergence of vacua is kept in tact.  Another possibility  is that the attractor becomes "softer", meaning that  the brane charge cannot decrease below a certain minimal value.   In the latter case, the scanning of the Higgs VEV cannot get  finer beyond the certain minimal step, and correspondingly the divergence in the number density of vacua gets cut-off at some maximal value.  We shall now give a more detailed discussion of the above two regimes. 

\subsection{A Positive Mass Term:  Sharpening the Attractor} 

 Consider the effect of the positive mass term first. Such a mass term, "repels" the $\phi$ field 
from the brane, and effectively diminishes its VEV at the brane location.  Ignoring the effect of $\sim q_{eff}$-terms, the 
equation for $\phi$ in the  background of the brane located at $z=0$ is 
\begin{equation}
\label{eqbrane}
\partial^2 \phi\, - \, ( \, m^2_{bulk}\, - \, \delta(z)M_B)\phi \, +\, \lambda \phi^3\, = \,0  
\end{equation}
where $m^2_{bulk}$ is the effective bulk mass term which includes the contribution coming from $F$
\begin{equation}
\label{bulkmass}
m_{bulk}^2 \, =\, m^2 \, + \, F_0^2/2M^2  
\end{equation}
 From (\ref{eqbrane})  it is clear that the positive brane-localized mass term is seen by 
the field as a potential barries, and for $M_B \gg m_{bulk}$ the expectation value at the brane location
$\phi(0)$ is considerably smaller than its bulk counterpart $\phi(\infty)$.  

 $\phi(0)$ can be estimated by minimizing the following expression
(we ignore the factors of order one)
\begin{equation}
\label{energymin}
E \, =\, M_B\phi(0)^2 \, + \, (\phi(0) - m_{bulk})^2m_{bulk}\, + \, (\phi(0)^2\, - \, m_{bulk}^2)^2m_{bulk}^{-1}
\end{equation}
The first term in this expression comes from the brane mass term. The second and the third terms are
the expenses in the gradient and the bulk potential energies.
The full expression is minimized at 
\begin{equation}
\label{p0min}
\phi(0) \sim {m_{bulk}^2 \over M_B}
\end{equation}
Thus, in any given vacuum, the brane expectation value of $\phi$ is 
by the factor ${m_{bulk}\over M_B}$ smaller then its bulk counterpart $\phi(\infty) \sim m_{bulk}$.
Thus, for a given value of the bulk mass term, the value of the brane charge $q_{eff}$ is by a factor
of $(m_{bulk}/M_B)^{N}$ smaller, as compared to what it would be for $M_B=0$. 
Correspondingly according to the general rule of vacuum counting, the number of vacua with 
a given VEV of the Higgs field $\phi \sim \phi_0$, becomes 
\begin{equation}
\label{newnvacua}
n_{\phi_0} \sim \left ( {M_P^N \over \mu \phi_0^{N-2}} \right )\left({M_B\over \phi_0}\right)^{N}, 
\end{equation}  
and is by factor of $\left({M_B\over \phi_0}\right)^{N}$ bigger than what it would be 
for $M_B\, =\, 0$.
Thus, the positive sign brane-localized mass term makes the attractor stronger. 

\subsection{A Negative Mass: Smoothing Out  the Attractor}

 Now let us show that in case of the negative sign the brane-localized mass term has an opposite effect,
and may either shift the attractor or cutt off the divergence of the number of vacua. 
This is because for the negative brane mass term, $\phi$ develops a non-zero expectation value
on the brane even for $m^2_{bulk}=0$ (that is, when the bulk VEV is zero).  Thus, the brane continues 
to have a non-zero charge $q$ even in  $\phi = 0$ vacua, and the step of change stays finite. 
As a result number of vacua gets cut-off.  

 Let us show that  for the negative sign of the brane mass term  $\phi$  indeed develops a non-zero value on the brane  in the limit $m^2_{bulk} \rightarrow 0$. 
The fact that $\phi$ wants to condense on the brane can be seen by examining the linearized equation for small perturbations about the $\phi = 0$ 
solution in the brane background. This equation has the following form 
\begin{equation}
\label{inst}
\left (\partial^2 \,  - \delta(z) M_B  \right ) \phi \, = \, 0 
\end{equation}
It is obvious that there is a  normalizable exponentially-growing tachyonic  mode, localized on the brane
\begin{equation}
\label{tachyoin}
\phi \, = \, e^{{1\over 2} M_Bt} e^{-{1\over 2}|z|M_B}
\end{equation}
This instability signals that $\phi$ condenses on the brane and develops a non-zero expectation value
there. This condensate is  $\phi(0) \sim M_B$,  since for $m_{bulk} =0$,  $M_B$ is the only mass scale in the problem.
 Hence, the expectation value on the brane is
$\sim M_B$ even though  the bulk VEV vanishes.

 This fact has profound implications for the attractor dynamics, since now even in the vacua with 
$m_{bulk}\, = \, 0$ the brane charge is non-zero  and is given by
\begin{equation}
\label{qmin}
q_{min} \, \sim \, \mu (M_B/M_P)^N
\end{equation}
To see what these implications are, first note 
that the value of $F_0$ in the vacuum with  $m_{bulk} \, =\,  0$ is $F_0^2\, = \, 2M^2m^2$.  Then, 
if 
\begin{equation}
\label{mbranegg}
M_B^2 \gg 4 \sqrt{2}{m\over M} q_{min},
\end{equation} 
the attractor will be shifted to a positive value $m_{bulk}^2 \, =\, M^2_B/4$.  In the opposite limit
\begin{equation}
\label{mbranel}
M_B^2 \,< \,4 \sqrt{2}{m\over M}q_{min}
\end{equation}
the attractor will be regulated, and the divergence in the vacuum number density
will be cut-off at 
\begin{equation}
\label{nmax}
n_{max} \sim {mM \over q_{min}} 
\end{equation}
The above two regimes can be understood from the fact that  the equation
\begin{equation}
\label{rhopert}
(\partial^2 \, - \, m_{bulk}^2 \, - \, \delta(z) M_B)\phi \, = \, 0 
\end{equation}
has an exponentially growing normalizable tachyonic mode
\begin{equation}
\label{tachyoin}
\phi \, = \, e^{{1\over 2} t\sqrt{M_B^2+4m^2_{bulk}}} e^{-{1\over 2}|z|M_B},
\end{equation}
as long as
 \footnote{For $M_B^2 \,+\, 4m_{bulk}^2\,=\,0$ there is a localized zero mode
$\phi \, = \, \beta {\rm e}^{-{|z|M_B \over 2}}$
that can be given an arbitrary expectation value without costing any energy up to bilinear in $\phi$.
In the other words, for $\lambda = 0$ there is a one-parameter class of the zero energy
solutions with arbitrary $\beta$.}
\begin{equation}
\label{condtach}
M_B^2 \,+\, 4m_{bulk}^2>0.
\end{equation}
Thus, only in this regime $\phi$ develops a non-zero VEV in the vicinity
of the brane. If (\ref{mbranel}) holds in $m_{bulk}=0$ vacuum, then the condition (\ref{condtach})
will be violated within a single step, and the branes will become chargeless.  Thus, $q_{min}$ is the smallest possible non-zero brane charge. Due to this the singularity in the  number density of vacua will get
smooth-out at this point, according to (\ref{nmax}).

\section{Realistic model building and predictions}.

\subsection{The Need of a Second $SU(2)\times U(1)$-Doublet.}

 So far we have been discussing the attractor solution of the hierarchy problem on a toy example
in which the prototype for the Standard Model Higgs was a complex singlet $\phi$. 
In order to implement this  idea in the realistic model, we have to promote $\phi$ into
the doublet representation of  $SU(2)\times U(1)$ group.  This creates an issue of how to write down the 
gauge invariant interaction with the brane. Since $C_{\alpha\beta\gamma}$ carries no electroweak quantum numbers, 
$q(\phi)$ must be an $SU(2)\times U(1)$-invariant, but $Z_{2N}$-{\it odd}  function of $\phi$.  This is, however, impossible to achieve by employing a single Higgs 
doublet, since the only possible  gauge invariant $\phi^*\phi$ is also automatically an $Z_{2N}$-invariant. This in unacceptable, since $Z_{2N}$-{\it oddness}  of the brane charge is what guarantees
the UV-stability of the attractor point. 
 Fortunately, there are number of ways for circumventing the above technicality. Here we shall
discuss one of the simplest and economic ones.  
%Some alternatives will be discussed in Appendix F. 

\subsection{Quark condensate as a second doublet} 

The QCD condensate of the light quarks in the Standard Model carries the quantum  numbers identical 
to the Higgs doublet and contributes to the Higgsing of $SU(2)\otimes U(1)$-symmetry. 
In fact, because of this quark condensate, the electroweak $W$ and $Z$ bosons would get masses even in the absence of the Higgs scalar\cite{lennie1}. 
We can use this condensate in convolution with  the Higgs  doublet  for creating  the gauge-invariant  (but $Z_{2N}$-{\it odd}) brane charge.  There are two lowest possible gauge invariants
(per generation)  that can be
made out  of the Higgs and the quark bilinears. These are ($SU(2)$-indexes are suppressed)
\begin{equation}
\label{invup}
\phi \bar{Q}_LU_R
\end{equation}
and 
\begin{equation}
\label{invdown}
\phi^*\epsilon \bar{Q}_LD_R
\end{equation}
respectively. Here $Q_L$ is the left-handed quark doublet, and $U_R,D_R$ are the right-handed up
and dawn quarks respectively.  $\epsilon$ is an antisymmetric $SU(2)$ tensor.  Due to QCD quark condensate, the expectation value of these invariants is non-zero as long as $\phi\neq 0$. 
Thus, any of these invariants can be used for creating the attractor at $\phi \, = \, 0$.
 For this we have to require that the given invariant, e.g., (\ref{invup})
 (or any power of it) is odd under the brane 
conjugation symmetry. For example, 
\begin{equation}
\label{znonquarks}
(\phi \bar{Q}_LU_R)  \, \rightarrow \, e^{i{\pi\over N}} (\phi \bar{Q}_LU_R)
\end{equation}
Then the brane charge becomes the function
of (\ref{invup})
\begin{equation}
\label{qup}
q \, = \, {\mu\over 6} Re\left ({ \phi \bar{Q}_LU_R \over M_P^4 }\right )^N\,
\end{equation}
Note that since (\ref{invup}) transforms non-trivially under $Z_{2N}$-conjugation, it cannot 
appear in the Lagrangian. Thus, the  Standard Model Yukawa coupling  that could give
a diagonal mass to the given
$U$-quark is forbidden in this theory.  This fact highlights a generic feature of the 
model, irrespective which quark condensate creates a brane charge,  appearance of 
certain zeros in Yukawa matrix is the generic prediction.
Precise structure depends on the $Z_{2N}$  charge assignment, and will not be discussed here, 
but it is an important aspect for understanding the predictivity of the attractor solution.   

\subsection{A complete model}

Putting  all the ingredients together we can now write down a simple extension of the 
Standard Model which solves  the hierarchy problem via the attractor mechanism. 
The action is
\begin{eqnarray}
\label{complete}
S \, &=& \, \int_{3+1} \, |D_{\mu}\phi|^2 \,  - \,  {1 \over 48} F^2 \, + \,  \, |\phi|^2 \, \left (m^2\,  + \,  {F^2 \over 48M^2}
\right) \, - \, {\lambda \over 2} \, |\phi|^4 \, + \\ \nonumber
&& -{1\over 6} \, C_{\alpha\beta\gamma} J^{\alpha\beta\gamma}_{(T)} \, 
% {J_{\mu} \over 6} \Theta^{\mu\nu} \epsilon_{\nu\alpha\beta\gamma} C^{\alpha\beta\gamma} \,
-  M_{br} |\phi|^2,
\end{eqnarray}
plus the usual action of the Standard Model.  In (\ref{complete}) $D_{\mu}$ is the covariant derivative. 
$M_{br}$ is the brane-localized mass term given by (\ref{branemassterm}), and we choose the sign to be positive.
The current $J^{\alpha\beta\gamma}_{(T)}$ is given by (\ref{current}) with  
  \begin{equation}
\label{qhphi}
q_{eff} \, = \,  {\mu\over 12} \left[ \left ({ \phi \bar{Q}_LU_R \over M_P^4 }\right )^N \, - \,
\left ( {\bar{Q}_LU_R \bar{Q}_LD_R \over M_P^6 } \right )^K \right ] \, + \, h.c.
\end{equation}
 In this model we choose to use the quark condensate 
as a second Higgs doublet in order to construct a gauge-invariant brane charge. 
The attractor value is shifted away from $\phi=0$ by the second quark condensate in the 
figure brackets.  For the latter invariant to be non-zero the quarks must be taken from different generations. 
The integers $N$ and $K$ are determined by the transformation properties of the various fields  under the  discrete symmetry.  This transformation properties also restrict the structure of Yukawa 
matrix elements. For instance, as said above for arbitrary $N$, the diagonal Yukawa coupling of the 
up quark is forbidden.  

 Taking all these terms into the account the attractor value for the bulk Higgs VEV is
\begin{equation}
\label{attrvev}
\phi_{attr} \, \sim \, (M_B^{{1\over 2}}M_P^{{2\over 3}}) \left ({\Lambda_{QCD} \over M_P}
\right )^{\left (4{K\over N} -2 \right )}
\end{equation}
Where $\Lambda_{QCD} \sim $GeV is the strong interaction scale. 
 For $M_B \sim M_P \sim 10^{19}$GeV,  The correct attractor value is established 
around $K /N\simeq 5/7$ or so.
Since the brane charge-conjugation symmetry implies nontrivial restrictions on the matrix of Yukawa
couplings, it would be interesting to classify predictions of fermion mass relations for various assignments that lead to the 
correct attractor value. 

 We wish to notice that other Standard Model parameters, such as, for example,  Yukawa coupling constants,  can (and in general will) depend on the values of $F$ and $\phi$ through some high-dimensional 
$M_P$-suppressed operators. The question then is how the attractor  influences the values
of such parameters.  To answer this question it is useful to classify  parameters by their transformation properties 
under $Z_{2N}$.  The general rule then is that the parameters that are $Z_{2N}$-even 
do not change significantly near the attractor point, since in the zeroth order such parameters do not depend  on the $\phi$-VEV.  The example of $Z_{2N}$-even parameters are, for instance, all the Yukawa coupling constants of the couplings that are allowed 
by the $Z_{2N}$ symmetry. The dependence of such a coupling constant on  $\phi$ and $F$ can be 
parameterized  in form of the expansion  in series of invariants $F^2/M_P^4$ and $|\phi|^2/M_P^2$
\begin{equation}
\label{yukawaexp}
g\, = \, g_0 \, + \, g_1F^2/M_P^4 \, + \, g_2| \phi|^2/M_P^2 \, + \,...  
\end{equation}
where $g_0, g_1, g_2$ are field-independent constants. It is obvious that unless $g_0$ is minuscule, the change  of $g$ near the attractor point is negligible, since both $\Delta F$ and $\Delta \phi$ vanish there. So generically,  at the attractor point the expectation values of  $Z_{2N}$-even parameters will be set by some attractor-insensitive physics.

 \subsection{The Heavy Higgs Doublet}

An alternative to the quark condensate for creating and $SU(2)\times U(1)$-invariant brane charge,  is to introduce a second Higgs doublet $H$.  We shall assume that 
$H$ has a positive mass square $M_{H}^2$ of order $M_P^2$, and has no expectation value in the 
bulk vacuum.  However,  $H$ is allowed by symmetries to have a large brane-localized mass term. 
We shall choose the sign of this mass term to be negative  so that $H$ develops a non-zero VEV
on the brane.  This is sufficient for forming the $Z_{2N}$-odd but  $SU(2)\times U(1)$-invariant
brane charge out of the two doublets.  

To achieve this we shall require that the $SU(2)\times U(1)$-invariant product of the two doublets
transforms under the brane "charge conjugation" $Z_{2N}$ symmetry
\begin{equation}
\label{z2}
(H^*\phi) \, \rightarrow \, e^{i{\pi\over N}} (H^*\phi)
\end{equation}
 The new $H$-dependent terms in the action are 
\begin{equation}
\label{hfp}
S\, = \, \int_{3+1} \, -{1\over 6}\, C_{\alpha\beta\gamma} J^{\alpha\beta\gamma}_{(T)} \, + \, 
M_{br}(x)^2|H|^2\, + 
\, |D_{\mu}H|^2 \, -  \,M_H^2 \, |H|^2\, - \, {1\over 4} |H|^4\, + ... 
\end{equation}
The current $J_{(T)}$ is given by (\ref{transverse}) with  
  \begin{equation}
\label{qhphi}
q_{eff} \, = \, {\mu\over 6}\, Re\left ({ H^*\phi \over M_P^2 }\right ) ^N 
\end{equation}
The second term is the brane-localized mass term
\begin{equation}
\label{branemassterm}
M_{br}^2(x)\, =\, \int\, d\xi^3\sqrt{-g}\, M_{BH} \delta^4(x-Y),
\end{equation}
where $M_{BH} \, > \, 0$. 
In addition the total action will contain
all possible $SU(2)\times U(1)\times Z_{2N}$-invariant couplings among $H,\phi$ and $F$. 
These are unessential and are not shown for simplicity. The only requirement on such
cross-interaction terms of the form $|H|^2|\phi|^2, ~|H|^2F^2$ is that they do not  make the effective 
bulk mass$^2$ of $H$ negative. This is easy to arrange by choosing  the signs of these interactions, and can be assumed to be the case without any loss of generality. 
 Note that $Z_{2N}$ symmetry forbids the appearance of the mixing term $H^*\phi$ in the action.

 As long as 
\begin{equation}
\label{mhcond}
 M_{BH}^2 \, > \, 4M_H^2,  
\end{equation}
$H$ develops a non-zero VEV on the brane. 
This is because for (\ref{mhcond})  the linearized equation for $H$ in  the brane background
(located at $z=0$)
\begin{equation}
\label{heq}
(\partial^2 \, + \, M_H^2\, -\,M_{BH} \delta(z))H\, = \, 0
\end{equation}
has a localized exponentially-growing tachyonic mode 
\begin{equation}
\label{hsolution}
H \, = \, {\rm e}^{{1\over 2} t\sqrt{M_{BH}^2 - 4M_H^2}} {\rm e}^{-{1\over 2} M_{BH}|z|}.
\end{equation}
Because $H$ is non-zero on the brane,  the effective brane charge  (\ref{qhphi})
vanishes only for $\phi \rightarrow 0$, and vacuum $\phi=0$ is an attractor.

\section{The doublet-triplet splitting}

 In Grand Unified Theories  (GUTs) the low energy supersymmetry alone cannot guarantee
the smallness  of the Higgs mass, due to the problem of Doublet-Triplet splitting. The problem can be illustrated on an example of a simplest  $SU(5)$-extensions of the standard model. 
Because of  $SU(5)$ symmetry the Higgs doublet $\phi$ acquires a color-triplet   
partner ($T$), and the two together form the five-dimensional representation of the $SU(5)$ group,
$5_{Higgs} \, = \, ( \phi, T)$. Thus,  because of GUT symmetry both the weak doublet and 
the color-triplet  are forced to couple to quarks and leptons, that transform as $10\,  + \, \bar{5}$ dimensional 
representations per generation
\begin{equation}
\label{fivecouplings}
5_{Higgs} \, 10 \, 10\,  +  \bar{5}_{Higgs}^* 10 \bar{5} .
\end{equation}  
$SU(3)\times SU(2)\times U(1)$-reduction of the above coupling shows that the three-level 
exchange of the color-triplet  Higgs violates the baryon number and would mediate an unacceptably
fast proton decay, unless $T$ acquires a very large mass due to GUT symmetry breaking. 
In the same time $\phi$ should stay light, and this requires an additional fine tuning not 
provided by supersymmetry alone.  Let  $\Sigma_{i}^{j}$ be an $SU(5)$-adjoint Higgs that breaks 
the GUT symmetry ($i,j=1,2,...5$ are $SU(5)$-indexes).   The doublet-triplet mass splitting 
is accomplished through the coupling of $5_{Higgs}$ with $\Sigma$
\begin{equation}
\label{sigmacouplings}
5^*_{Higgs} (a \Sigma^ 2 \, + \,  b\Sigma) 5_{Higgs}  \, + \, (c {\rm Tr} \Sigma^2\, +\, m^2)  5^*_{Higgs} 5_{Higgs} 
\end{equation}
where $a,b,c,$ are some constants. 
After $\Sigma$ develops the VEV $\Sigma \, = \, {\rm diag} (2, 2,2,-3,-3)\sigma$ the masses of 
$\phi$ and $T$ become split as 
 \begin{equation}
\label{doubletmass}
m_{\phi}^2 \, = \,  (9a \, + \, 30 c) \sigma^2 \, - \, 3b\sigma   \, + \, m^2
\end{equation}
and
\begin{equation}
\label{tripletmass}
m_{T}^2 \, = \,  (4a \, + \, 30 c) \sigma^2 \, + \, 2b\sigma   \, + \, m^2
\end{equation}
respectively.  
The additional fine tuning amounts to setting (\ref{doubletmass}) to $\sim (100{\rm GeV})^2$. 

 We wish to point out now that the attractor  solution of the Hierarchy Problem,  automatically  solves the problem of doublet-triplet mass splitting in Grand Unified Theories. 
 Applying the attractor idea to GUTs simply amounts to promoting 
$m^2$ in (\ref{doubletmass}) and (\ref{tripletmass}) into the function of 
the four-form field-strength $F$.  Then, just as in case of Standard Model, $m_{\phi}^2$ is attracted 
to a small value, and in the same time $m_{T}^2$ is attracted to a large one. 
 Note that depending of the parameters, because of $SU(5)$ symmetry of the brane charge, there may be other attractor points, e.g.,   at $m_{T}^2 \, = \, 0$. 

\section{Discussions and Outlook}

 The  attractor solution is probably as close as the axion-type dynamical relaxation mechanism 
could come to the solution of the hierarchy problem. Indeed, the first step of the attractor 
solution is exactly to bring the hierarchy problem on the same footing as the Strong CP 
Problem\cite{theta} in QCD.   The latter, as we know, is not the problem  of UV-sensitivity 
but rather the problem of the super-selection rule among the infinitely many $\theta$-vacua. 
As we have shown in Section 3.1, this is precisely what happens to the Hierarchy Problem
when we couple Higgs to the three-form field:   
From the problem of UV-sensitivity of the Higgs mass, it gets converted into a 
{\it super-selection problem} of the latter.  The continuum of the vacua scanned by  
$m_{\phi}$ are all good, very much like QCD $\theta$-vacua in QCD.  

Having achieved this, we realize that there is a profound difference between the problem of UV-sensitivity and the problem of super-selection. The solution of UV-sensitivity problem always 
requires a new strongly-coupled  physics at low energies, whereas the solution of the super-selection problem  does not.  In the former case the new strongly-coupled physics is required in order to regulate  the quadratic divergency in the Higgs mass. An example for such a regulating physics is the law energy  supersymmetry.  Whereas, the new physics which solves the super-selection problem, can be {\it arbitrarily weakly} coupled. The good example of such new physics in case of the Strong CP problem, is the axion,  which can have an arbitrarily high scale and be practically invisible.  
 Like-wise, in our case the new physics that selected the attractor vacuum  is arbitrarily decoupled, 
and can be  practically unobservable at low energies.  Nevertheless, as we have seen already the minimal realistic models can be predictive and potentially testable.   Attractor  are the mechanism  through which the multi-vacua fundamental theory could make sharp low energy prediction. 

 If the attractor solution of the Hierarchy Problem  can be fully implemented in String Theory, it will most likely  have "softer" properties. For instance, the infinite number density of vacua probably will be regulated,  as already suggested by the distribution of vacua in\cite{md, shamit}. 
This should not be an obstacle for solving the hierarchy problem, provided the number of vacua is sharply peaked around the small Higgs mass.  
Moreover, as we so in section 5.2,  the softened attractors can appear already in effective field theory treatment. 

  In theories with attractor vacua  there are number of open important questions, such as, whether 
some version of the attractor mechanism could select the small cosmological constant. 
Some of these questions will be addressed in \cite{giga}

\vspace{0.5cm}   

{\bf Acknowledgments}
\vspace{0.1cm} \\

 It is pleasure to thank  Andrei Linde, Alex Vilenkin and 
especially Gregory Gabadadze and Shamit Kachru for useful discussions and comments.
We are grateful to Neal Weiner for raising interesting questions on cosmology 
of attractors, and for discussions.   
We also thank Michael Douglas  for comments. 
This work is supported in part  by David and Lucile  Packard Foundation Fellowship for  Science and Engineering, and by NSF grant  PHY-0245068

\section{Appendix}

\subsection{Appendix A:  Gauge Invariance from  Goldstones} 

 In this appendix we shall discuss how the coupling (\ref{newcoupling}) can be obtained by  integration
of Goldstone-type degrees of freedom in  a local,  gauge-invariant theory.  Let us first illustrate the idea
on an example of electrodynamics. Imagine, that we wish to couple a photon $A_{\mu}$ to 
a non-conserved current 
$J_{\mu}$ ($\partial_{\mu}J^{\mu} \neq 0)$ in a gauge invariant way. For example, such can be 
a $1+1$-dimensional version of the current (\ref{current})
\begin{equation}
\label{current1}
J^{\mu}(x)\,  = \,  \int d\xi \delta^2(x \, - \, Y(\xi))\, q(Y)\,  
 \left( {\partial Y^{\mu} \over \partial \xi} \right) 
\end{equation} 
which in general  is not conserved unless  $q$ is a constant.  The conservation can be restored 
by introducing additional degrees of freedom, that will compensate the divergence of 
(\ref{current1}) for non-constant $q$. 

This can be accomplished by introducing a compensating St\"uckelberg field $\theta$. The coupling to the
source then can be written in the following form 
\begin{equation}
\label{thetacoupling}
(A_{\mu} \, - \, \partial_{\mu}\theta)\, J^{\mu}
\end{equation}
This coupling will be gauge invariant if we demand that under the gauge transformation
\begin{equation}
\label{gaugetr}
A_{\mu} \, \rightarrow \, A_{\mu} \, + \, \partial_{\mu} \omega
\end{equation}
the compensating field shifts as 
\begin{equation}
\label{thetashift}
\theta \, \rightarrow \, \theta \, + \, \omega 
\end{equation}
Hence, $\theta$ is in fact a Goldstone field.  Because of the gauge invariance, the Lagrangian can
only depend on $\theta$ through the combination $(A_{\mu} \, - \, \partial_{\mu}\theta)$. 
Since current $J_{\mu}$ is not conserved, to maintain the gauge invariance we have to impose 
an additional constraint.
\begin{equation}
\label{thetaconst}
\partial^2 \theta \, = \, \partial_{\mu}A^{\mu} 
\end{equation}
Then, integrating out $\theta$ we can
write down an effective Lagrangian for $A_{\mu}$
\begin{equation}
\label{photoneffective}
L \, = \, - {1\over 4} F_{\mu\nu}F^{\mu\nu} \, + \, J^{\mu} \Pi_{\mu\nu}A^{\nu}\,
\end{equation}
We see that after integrating out of the St\"uckelberg field, photon only couples to the transverse
part of $J_{\mu}$, as required by gauge invariance. 

 The generalization of the above construction to the case of the three-form field $C_{\alpha\beta\gamma}$ is straightforward.  Again in order to achieve a gauge-invariant  coupling of the three-form field to a non-conserved current $J_{\alpha\beta\gamma}$, we introduce a 
compensating two-form $B_{\alpha\beta}$, which under the gauge transformation (\ref{gauge})
shifts in the following way
\begin{equation}
\label{gaugeB}
B_{\alpha\beta} \, \rightarrow  \, B_{\alpha\beta}  \, + \, \Omega_{\alpha\beta},
\end{equation}
The gauge invariant coupling to an arbitrary non-conserved  source $J_{\alpha\beta\gamma}$ is
\begin{equation}
\label{thetacoupling}
(C_{\alpha\beta\gamma} \, - \, F_{\alpha\beta\gamma}^{B})\, J^{\alpha\beta\gamma},
\end{equation}
where $F^{B}$ is the field strength of $B$
\begin{equation}
\label{FB}
F_{\alpha\beta\gamma}^B\, = \, d_{[\alpha} B_{\beta\gamma]}
\end{equation}
%The action is 
%\begin{equation}
%\label{massiveC}
%L \, = \, - {1\over 48} F^2 \, + \, {m^2 \over 12} (C_{\alpha\beta\gamma} \, - \, %F_{\alpha\beta\gamma}^{B})^2 \, + \, g (C_{\alpha\beta\gamma} \, - \, F_{\alpha\beta\gamma}^{B})\, %J^{\alpha\beta\gamma},
%\end{equation}
As in the photon example, we have to impose the following  constraint on $B$ 
\begin{equation}
\label{Beq}
\partial^{\mu} F^B_{\mu\alpha\beta} \, = \, 
\partial^{\nu} C_{\nu\alpha\beta} \, 
\end{equation}
This constraint can be enforced by introducing an additional auxiliary two-form field $X^{\beta\gamma}$
with the following coupling in the Lagrangian
\begin{equation}
\label{xcoupling}
X^{\beta\gamma} \partial^{\alpha} (C_{\alpha\beta\gamma} \, - \, F_{\alpha\beta\gamma}^{B}).
\end{equation}
The equation of motion of $X^{\alpha\beta}$ then imposes the constraint (\ref{Beq}).
Now integrating out the $B_{\alpha\beta}$-field we write down the effective Lagrangian for $C$
\begin{equation}
\label{CBeffective}
L \, = \, - {1\over 4} F_{\mu\nu}F^{\mu\nu} \, + \, {1\over 6} J^{\mu\alpha\beta} \Pi_{[\mu}^{\nu}
C_{\nu\alpha\beta]}\,
\end{equation}
The above effective coupling coincides with (\ref{newcoupling}) up to a total derivative. 

\subsection{Appendix B:  The Charge Screening}

 We shall now discuss a potential effect,  which may lead to the screening of the 3-form charge 
by the brane loops. The effect is somewhat similar in spirit to the charge screening by fermion 
loops in the massless  Schwinger model\cite{screening}, except that this issue in our case is more subtle, as we shall
now discuss.  For simplicity let us consider the analogous question in 
electrodynamics first.  Consider the action given in (\ref{photoneffective}) 
where $J_{\mu}$ is some generic current. Since the photon couples only to its transverse part, the 
theory is automatically gauge invariant regardless whether the current  $J_{\mu}$ is conserved or not. 
Consider now a correlator of the two currents
\begin{equation}
\label{jjcor}
\langle J_{\mu}(x) J_{\nu}(x') \rangle \, = \, P \Pi_{\mu\nu} \, + R \,\Theta_{\mu\nu} 
\end{equation}
where $P$ and $R$ are some scalar functions of the cut-off and the momentum.  If the transverse part of the correlator is nonzero ($P \neq 0$),  it 
will generate the following effective operator in the photon action
\begin{equation}
\label{effectiveP}
A^{\mu} P\Pi_{\mu\nu}A^{\nu} \,
\end{equation}
Depending on the structure of the operator $P$ this can be interpreted as the correction either to the 
photon kinetic terms, to the mass term, or to both.   In case, if the propagator 
${1\over \partial^2 + P}$ has a physical pole, the (\ref{effectiveP}) generates the mass for photon, 
and charges will be screened.  This is in particular the case for  $P =$constant.  
The result depends on the underlying structure of the theory. For instance, for the conserved
fermionic  current  $J^{\mu}$,
(\ref{photoneffective}) is equivalent to usual massless electrodynamics, which gives different 
answers in different number of dimensions.  In  $3+1$ dimensions, for the conserved fermionic current 
$P \propto \partial^2$, and no photon mass is generated.  On the other hand in $1+1$ Schwinger model
the answer depends on the fermion mass\cite{screening}.  In case of massless fermions, the photon mass is generated and the charges are screened.  For massive fermions, however, the screening is only partial.  

 Let us now show that the analogous question can be posed in our case.  Consider the coupling
(\ref{newcoupling}). The correletor of the two currents $J_{\mu}$ can again be parameterized in the
form (\ref{jjcor}).   If $R\neq 0$ this will generate the following operator
\begin{equation}
\label{CR}
%\epsilon_{\mu\alpha\beta\gamma} C^{\alpha\beta\gamma}  
C^{\mu\alpha\beta} \, R \,\Pi_{[\mu}^{\nu}
C_{\nu\alpha\beta]}
%(R\Theta^{\mu\nu})C^{\tau\rho\kappa} \epsilon_{\nu\tau\rho\kappa}
\end{equation}
which would signal the generation of mass, if ${1\over (1 + R)}$ had a physical pole. 
%This is also obvious, if we note that for the constant $R \, = \, m^2$,  (\ref{CR}) is equivalent  to 
%the second term in (\ref{CBeffective}). 
 Because in general $R$ is expected to be a function 
of $\partial^2/M^2$, the requirement of the absence of the physical poles below the cut-off scale
reduces to the requirement of the absence of the constant part in  $R$.  In our case, $R$ depends 
on the loops of the super-heavy branes, and will not be attempted to be calculated.

\subsection{Appendix C: Exact Solutions}

 The effective equation eq(\ref{pot}) was derived in the approximation of constant $\phi$ per unit step.
In reality, there will be a small back reaction on $\phi$ in each individual step, which will lead to 
the re-adjustment of the brane charge, and subsequent re-adjustment of the VEVs in the bulk. 
It is obvious that in the attractor neighborhood, this back reaction is negligible, but it is instructive to 
take it into account for the completeness.  

 So we shall now derive the effective bulk equation determining the VEV of $\phi$ without 
ignoring the variation of $\phi$ in an elementary step. 
For this, we shall solve the equation (\ref{firstgrad}) and (\ref{secondgrad}) 
without the simplifying approximations  listed below them.
Substituting the form (\ref{solution}), the  eq(\ref{firstgrad}) now becomes
\begin{equation}
\label{firstgrad1}
\partial^{\nu} \left ( ( 1 -  |\phi|^2/M^2) F_0\right) \, =
 \, -\mu \Theta^{\nu z} \left [Re \left ({\phi\over M_P}\right ) ^N \, \delta(z)
\right ] 
\end{equation}
or equivalently 
\begin{equation}
\label{firstgrad2}
 \partial^2 \left (( 1 -  |\phi|^2/M^2) F_0\right ) \, =
 \, -\mu \partial_ z \left [Re \left ({\phi\over M_P}\right ) ^N \, \delta(z)
\right ] 
\end{equation}
The solution with the correct boundary conditions is
\begin{eqnarray}
\label{firstgradsol}
F_0 (x_j,z) \, = \,  {\mu  \over 2M_P^N \left ( 1 -  {|\phi(x_j, z)|^2\over M^2} \right )} \,
\big [Re (\phi^N (z=0, f_i(x_j, z))) \theta(z)  \, - \\ \nonumber
\, Re (\phi^N (z=0, f_i(x_j, -z))) \theta(-z) \, + \, f_0 (M_P^N/\mu)\big ]. 
\end{eqnarray}
where $\theta(z)$ is the step function, and  $x_i, ~i=0,1,2$ are the three remaining space-time coordinates 
parallel to the brane.  For clarity,  we have indicated the coordinate dependence, in order to 
stress how the values of the $\phi$-field at different locations determine the bulk value of 
$F_0$. 
The functions $f_i(z)$ are such that $f_i(x_j, z)|_{z=0} \, = \, x_i$,
\begin{equation}
\label{condbox}
\partial^2 \phi^N (z=0, f_i(x_j, z)))\, = \, 0
\end{equation}
%and 
%\begin{equation}
%\label{condder}
%\partial_z \phi^N (z=0, f_i(x_j, z)))|_{z=0} \, = \, 0
%\end{equation}
 Note that the function  
$\phi ^N(z=0, x_i)$ is the value of $\phi^N$ at the brane location.  
$f_0$ is the integration constant.  Thus, we see that for small $\phi$ the value of $F_0(z)$ in the bulk vacuum $z \neq 0$ is essentially determined by the value of the $\phi^N$ function {\it at the  brane location} $z=0$. For $z \neq 0$ the dependence of $F_0(z)$ on a local values of $\phi(z)$ is rather mild and has additional suppression factor 
$\phi^2/M_P^2$. 

 For instance, for the background values that are funbctions of $z$ and $t$ only, the solution is
\begin{eqnarray}
\label{fzt}
F_0 (z,t) \, = \,  {\mu  \over 2M_P^N \left ( 1 -  {|\phi(z, t)|^2\over M^2} \right )} \,
\big [(Re (\phi^N (0, (t-z))\, \theta(z)  \, - \\ \nonumber
\, Re (\phi^N(0,(t+z))) \theta(-z) \, + \, f_0 (M_P^N/\mu)\big]. 
\end{eqnarray}
 From here it is clear that the change of expectation value of $\phi$ on the brane triggers 
the corresponding change in $\phi$, which propagates away from the brane at the speed of light. 
Existence of such waves indicates the presence of some "hidden" massless degrees of freedom. 
This is not surprising, since the degree in question is the St\"uckelberg field $B_{\mu\nu}$, which we 
have integrated out.  In some sense, this degree of freedom in our case plays the role analogous to the
one  played by the  invisible axion in the solution of the Strong CP problem in QCD. 

Substituting the solution (\ref{fzt}) into the equation (\ref{secondgrad}), we get the following
bulk equation for $\phi$
\begin{eqnarray}
\label{secondgradeff}
\partial^2 \phi \, - \, \Big (\, m^2\, - \, 
 {\mu^2  \over 8M^2 M_P^{2N} \left( 1 -  {|\phi(x_j, z)|^2\over M^2} \right )^2} \,
\big [(Re (\phi^N (0, (t-z))\, \theta(z)  \, - \\ \nonumber
\, Re(\phi^N(0,(t+z))) \theta(-z)  \, + \, f_0(M_P^N/\mu) \big ]^2 \Big) \, \phi \, + \, \lambda\, |\phi|^2\phi
\,  = \, 0
\end{eqnarray}
 Since in the vacua  $F_0$ takes discrete values that between the 
two neighboring vacua are spaces by $\sim \phi^N$,  for small $\phi$ the scanning becomes almost contineous. So the dynamics of the small-$\phi$ vacua can be studied by considering the 
appropriate values of the integration constant $f_0$.  In small $\phi$ vacua this constant takes the values
\begin{equation}
\label{smallphiconstant}
f_0^2 \simeq  8m^2 M^2
\end{equation}
Expanding (\ref{secondgradeff}) about such a vacuum,  and noticing that the value of the 
$\phi^N$ determining $F_0$ is taken at the brane location, we see that left-interactions are weak 
in the attractor vacuum.

\subsection{Appendix D: Resolving the Brane}
 
We shall now resolve the structure of the brane, and show that it can arise in the effective low energy theory in form of  an "axionic" domain wall\footnote{The "axion"  
should not be 
confused with the one solving the Strong CP problem. The role of the former can be played by one of the axions appearing in string compactifications.}
 Such a possibility was suggested in \cite{dv1, dv}, but we shall review it here in more details for completeness.   For this we shall introduce an 
"axion" field $a$, defined modulo $2\pi$, with the decay constant $f_a$. 
First, let us show that in presence of the coupling between the axion and the three-form field $C_{\alpha\beta\gamma}$, the 
axionic domain walls become branes charged under $C$\footnote{At the level of the present discussion we shall treat  both $a$ and $C$ as "elementary" objects, without specifying their possible origin from the fundamental underlying theory\cite{shamit1}.}.

 For demonstrating this, the possible couplings to the Higgs field $\phi$ plays no role and we shall ignore the latter  for simplicity.  Consider then the 
following Lagrangian
\begin{equation}
\label{aF}
L \, =  {f_a^2 \over 2} \, (\partial_{\mu}a)^2 \, - \, V(a) \, - {q\over 12\pi} \, \partial_{\alpha} aC_{\beta\gamma\delta}\epsilon^{\alpha\beta\gamma\delta}\, -  \, {1 \over 48} F^2, 
\end{equation}
where $V(a)$ is the axion potential, which is required to be periodic under 
\begin{equation}
\label{ashift}
a\, \rightarrow \, a \, + \, 2\pi n
\end{equation}
The equations of motion are
\begin{equation}
\label{feq}
\partial^{\mu}  F_{\mu\nu\alpha\beta} \, = \, {q\over 2\pi } \, \partial^{\mu}  a \epsilon_{\mu\nu\alpha\beta}
\end{equation}
\begin{equation}
\label{aeq}
f_a^2\partial^2 a \, + \, V_a \,  - {q\over 12\pi} {F^{\mu\nu\alpha\beta} \over 24} \epsilon_{\mu\nu\alpha\beta}\, = \, 0
\end{equation}
Integrating the first equation, we get 
\begin{equation}
\label{fanda}
F_{\mu\nu\alpha\beta} \, = \, {q\over 2\pi } \, (a \, + \, 2\pi k) \epsilon_{\mu\nu\alpha\beta}
\end{equation}
where $k$ is an integer integration constant. Under shift symmetry (\ref{ashift}) $k$ changes as
\begin{equation}
\label{kshift}
k\, \rightarrow \, k \, + \, n
\end{equation}
The relation (\ref{fanda}) implies that any axionic domain wall, through which $a$ changes by 
$\Delta a$ effectively acquires a three-form charge 
\begin{equation}
\label{qa}
q_{eff}  \, = \, {q \Delta a \over 2\pi} 
\end{equation}
The value  of $\Delta a$ through the elementary wall can be found by substituting  the solution
(\ref{fanda}) in to eq(\ref{aeq}). This gives the following effective equation for the axion
\begin{equation}
\label{effa}
f_a^2\partial^2 a \, + \, V_a \,  + \, {q^2\over 24\pi^2} (a\, + \, 2\pi k) \, = \, 0
\end{equation}
The value of the $\Delta a$ through the elementary wall is determined by the distance between the
neighboring minima of the effective potential:
\begin{equation}
\label{effpot}
V_{eff} \, = \, V(a) \,  + \, {q^2\over 48\pi^2} (a\, + \, 2\pi k)^2
\end{equation}
For the small $q$ and fixed $k$, the number of local minima is roughly $\sim V(a)_{max}/q^2$,
where $V_{max}(a)$ is the maximal value of $V(a)$. 
 The elementary step is $\Delta a \simeq 2\pi$.  Thus, the branes acquire an effective three-form charge 
$q_{eff}\, = \, q$.  The above counting of vacua  is not surprising, since the model (\ref{aF}) can be viewed  as the simplest four-dimensional generalization  of the massive Schwinger model, in which there is an analogous counting of states\cite{coleman}.

 Restoration of  the $\phi$-dependence of the brane charge, can now be done in a straightforward
way, by promoting $q$ into the function of $\phi$ given by (\ref{charge}). The gauge-invariant 
Lagrangian has the following form
\begin{eqnarray}
\label{af1}
L \, = \,  {f_a^2 \over 2} \, (\partial_{\mu}a)^2 \, - \, V(a) \, - {q_{eff}(\phi)\over 12\pi}
 \partial_{\alpha}a \, \Theta^{\alpha}_{\mu} C_{\beta\gamma\delta}\epsilon^{\mu\beta\gamma\delta}\,\\ \nonumber -  \, {1 \over 48} F^2 
+ \, |\partial_{\mu}\phi|^2 \,  + \,  \, |\phi|^2 \, \left (m^2\,  + \,  {F^2\over 48M^2}
\right) \, - \, {\lambda \over 2} \, |\phi|^4
\, + \, ...
\end{eqnarray}
The equations of motions are 
\begin{equation}
\label{aeq1}
f_a^2\partial^2 a \, + \, V_a \,  - {q_{eff}(\phi)\over 12\pi}
 {F^{\mu\nu\alpha\beta}
\over 24} \epsilon_{\mu\nu\alpha\beta}\, 
+\, (q_{eff}-derivatives) \, = \, 0
\end{equation}
\begin{equation}
\label{feq1}
\partial^{\mu} \left ( ( 1 +  |\phi|^2/M^2) F_{\mu\nu\alpha\beta} \right) \, - \, {q_{eff}(\phi) \over 2\pi}\, \Theta^{\gamma\mu} \partial_{\gamma} a \, \epsilon_{\mu\nu\alpha\beta} \, + \, (q_{eff}-{\rm derivatives}) \,  = \, 0
\end{equation}
\begin{equation}
\label{peq1}
\partial^2 \phi \, + \, \left (-\, m^2\, - \, {F^2 \over 48M^2}\right ) \, \phi \, + \, \lambda\, |\phi|^2\phi\, 
+ \,  {\mu N \over 24\pi} {\phi^{*N-1} \over M_P^N}\, 
\partial_{\alpha}a \, \Theta^{\alpha}_{\mu} C_{\beta\gamma\delta}\epsilon^{\mu\beta\gamma\delta}\,
= \, 0
\end{equation}
and again ignoring the derivatives of $q_{eff}$ and the terms of order $\phi^2/M^2$ in 
the l.h.s. of eq(\ref{feq1}), and
integrating the equation for $C$, we get the 
following effective equation for $\phi$ -vacua
\begin{equation}
\label{pvacua}
  - (m^2 - (a\, - \, 2\pi k)^2{\mu^2 \over 4\pi^2M^2} (Re(\phi^N/M_P^N)^2) \phi \, + \, \lambda \, \phi^3\,  = \, 0,
\end{equation}
where $a$ is defined from the equation
\begin{equation}
\label{avacua}
 V_a \,  - {q_{eff}^2\over 24\pi^2} (a\, + \, k)\, = \, 0
\end{equation}
as discussed above, for small $q_{eff}$ the minima of the $\phi$ potential  are at $a\, = \, 2\pi n$, 
which means that near the attractor point the $\phi$-vacua are defined from the equation (\ref{pot}).

\subsection{Appendix E: Axion-three-form couplings from massless scalars.}

 We shall now discuss how the axionic domain walls with Higgs-dependent 
three-form charges can be generated  by integrating out  some intermediate scalar fields, and show that 
this is only possible  if the scalars in question are exactly massless. 
%\begin{equation}
%\label{qaf}
%{q_{eff}(\phi) \over 12\pi}\,\partial_{\alpha} %a\,C_{\beta\gamma\delta}\epsilon^{\alpha\beta\gamma\delta}
%\end{equation}
The idea is to start with the local coupling
\begin{equation}
\label{aJ}
q_{eff} \partial_{\mu} a J^{\mu},
\end{equation}
where $J^{\mu}$ is the gauge-invariant current, with the divergence
\begin{equation}
\label{diver}
\partial_{\mu} J^{\mu} \, =\, c\, F\, + \, ...
\end{equation} 
where $c$ is some constant.  Then "integrating out" $J^{\mu}$ we  write the efective coupling 
 between $a$ and $C$.
 In order to fulfill this program, we shall consider a toy $1+1$ dimensional example first. 
This is a $1+1$-dimensional electrodynamics coupled to two pseudoscalars, an axion $a$ and an additional pseudoscalar $\chi$.  The Lagrangian is
\begin{eqnarray}
\label{achi}
L \, = \,  {1 \over 2} \, (\partial_{\mu}a)^2 \,  +    {1 \over 2} \, (\partial_{\chi}a)^2 
+ \, q_{eff} \partial_{\mu} a \partial^{\mu}\chi \, - \,  V(a)\, -  \,\\ \nonumber  - \, V(\chi) \,  - 
\tau \partial_{\alpha} \chi C_{\beta}\epsilon^{\alpha\beta}\, -  \, {1 \over 4} F^2 
\end{eqnarray}
Here $C_{\alpha}$ is the electromagnetic vector potential, and $\tau$ is some constant. 
 $q_{eff}$, which is understood to be the function  of the Higgs VEV, will be treated as a constant for a moment.  The reason for us to consider the above example, is that  electric field in 
$1+1$ shares some properties with  a three-form in $3+1$ dimensions. 
Just like the latter, the non-interacting $1+1$ electric fields has no propagating degrees of freedom, 
but its value can change in the presence of charges, which play the role analogous to
2-branes in four-dimensions.  

 Our aim is to start from the local Lagrangian (\ref{achi}), which for arbitrary 
$q_{eff}$  is invariant under the axionic shift symmetry  (\ref{ashift}) as well as under the gauge symmetry 
\begin{equation}
\label{gauge1}
C_{\alpha}\, \rightarrow C_{\alpha} \, + \, \partial_{\alpha} \Omega,
\end{equation}
which is an $1+1$-dimensional analog of (\ref{gauge}).
However, $a$ and $C$ do not couple directly, but through the 
"intermediate" field $\chi$. In fact, $\partial_{\mu}a$ couples to the current $\partial_{\mu}\chi$
whose divergence  is set by $F$.  So we expect that after integrating out $\chi$ 
we arrive to ($1+1$-dimensional analog of) the coupling (\ref{aF}). 
This coupling should guarantee that axionic "walls", which in $1+1$-dimensions are 
just particles, acquire electric charges controlled  by $q_{eff}$.
We shall now check if this indeed is the case.  The equations of motion are
\begin{equation}
\label{a1}
\partial^2 a \, + \, q_{eff} \partial^2 \chi \, + \, V_a \, = \, 0
\end{equation}
\begin{equation}
\label{chi1}
\partial^2 \chi \, + \, q_{eff} \partial^2 a \, + \, V_{\chi} \, - \, {\tau \over 2} F_{\mu\nu}\epsilon^{\mu\nu}\,  = \, 0
\end{equation}
\begin{equation}
\label{f1}
\partial^{\mu} \, F_{\mu\nu}\, = \, \tau \, \partial^{\alpha}\epsilon_{\alpha\nu}\chi
\end{equation}
The last equation is solved by 
\begin{equation}
\label{f1sol}
F_{\mu\nu} \, =  \,\epsilon_{\mu\nu} \, \tau(\chi\, + \chi_0)  
\end{equation}
where $\chi_o$ is some constant.  Substituting this result in eq(\ref{chi1}), we get the following effective
equation fo $\chi$
\begin{equation}
\label{chi1}
\partial^2 \chi \, + \, q_{eff} \partial^2 a \, + \, V_{\chi} \, +  \, \tau^ 2 \chi\, + \tau^2\chi_0\, = \, 0
\end{equation}
For the special choice of
\begin{equation}
\label{vchi}
V_{\chi} \, = - \tau^ 2 \chi\, - \tau^2\chi_0\,
\end{equation}
 the system of equations is
solved by 
\begin{equation}
\label{chisol}
\chi\, = \, - \, q_{eff} a 
\end{equation}
where $a$ satisfies the equation
\begin{equation}
\label{a2 }
\partial^2 a \, + \,{1 \over 1 -  q_{eff}^2} \, V_a \, = \, 0
\end{equation}
Because of $2\pi$ periodicity of $V(a)$, the latter equation always has an axionic domain 
wall solution, through which $a$ changes by $2\pi$. According to (\ref{chisol}) the corresponding change in $\chi$ is $\Delta \chi \, = \, - q_{eff} 2\pi$ and according to (\ref{f1sol}) the change in 
$F$ is $\Delta F = -q_{eff} \tau 2\pi$. Thus, axionic walls indeed acquire an electric charge.  However, such a behavior is a peculiarity of the choice (\ref{vchi}). In fact for no other choice of $V_{\chi}$ may
axionic walls have a charge proportional to $q_{eff}$.  Indeed, the fields at infinity on both sides of the wall must assume the constants values satisfying
\begin{equation}
\label{achiinf}
V_a \, = 0
\end{equation}
and eq(\ref{vchi}).  Unless the latter is identically zero, the change of $\chi$ through the wall 
will be determined by the neighboring minima of (\ref{vchi}), which contains no reference 
to $q_{eff}$. Hence, change of $F$ will not be set by $q_{eff}$ either. 

 The reason of why the choice (\ref{vchi}) is the only possible one is easy to understand from the symmetry point of view.  For such a choice the Lagrangian  can be rewritten as  
\begin{eqnarray}
\label{achi}
L \, = \,  {1 \over 2} \, (\partial_{\mu}a)^2 \,  +    {1 \over 2} \, (\partial_{\chi}a)^2 
+ \, q_{eff} \partial_{\mu} a \partial^{\mu}\chi \, - \,  V(a)\, -  \,\\ \nonumber  - \,  
{1\over 4} (\tau \chi \,\epsilon_{\mu\nu} -  \,  F_{\mu\nu})^2 
\end{eqnarray}
which has an exact shift symmetry 
\begin{equation}
\label{shiftx}
C_{\mu} \, \rightarrow \, C_{\mu} \, + \, b\epsilon_{\mu\nu}x^{\nu},~~~\chi \rightarrow \chi \, -\, b/\tau  
\end{equation}
Due to this symmetry, after integrating out $F$, the action for $\chi$ cannot be anything 
other than the one of a scalar field with vanishing potential. 

The fact why in the model (\ref{achi})  the existence of the branes with electric charge $\propto q_{eff}$ requires a very  special choice of $V(\chi)$ can also be understood in the following way.  In the
limit $V(a) \, = \, V(\chi)\,  = \, 0$ there are the two continuous shift symmetries.   
\begin{equation}
\label{ashiftc}
a \, \rightarrow \, a \, + \, {\rm constant}
\end{equation}
 and 
\begin{equation}
\label{chishiftc}
\chi \, \rightarrow \, \chi \, + \, {\rm constant}
\end{equation}
The corresponding currents are
\begin{equation}
\label{acurrent}
J_a^{\mu}\, = \, \partial^{\mu}a \, + \, q_{eff} \partial^{\mu}\chi
\end{equation}
and 
\begin{equation}
\label{chicurrent}
J_{\chi}^{\mu}\, = \, \partial^{\mu}\chi \, + \, q_{eff} \partial^{\mu}a
\end{equation}
respectively. 
Only the second shift symmetry is "anomalous" and corresponding current is not conserved
\begin{equation}
\label{divchicurrent}
\partial^{\mu}J_{\chi}^{\mu}\, = \, {\tau \over 2} F_{\mu\nu}\epsilon^{\mu\nu}\, 
\end{equation}
Hence, $\chi$ is the only speudo-Goldstone boson whose shift is tied directly to $F$.  But shift of 
$\chi$ is determined by $V(\chi)$ which in general carries no information about $q_{eff}$.  Only in the limit (\ref{vchi}), when the "bare" mass of $\chi$ is exactly canceled by the "anomaly" contribution,  
shift symmetry  (\ref{chishiftc}) remains exact and shift in $\chi$  adjusts to the minimal step 
of $a$ suppressed by the small charge $q_{eff}$. 

 It is usefull to view the above effect in fermionic language. Indeed, for 
$V(\chi) \, \propto \, {\rm cos} 2\sqrt{\pi}\chi$, 
the Lagrangian
(\ref{achi}) is a bosonised version of the following fermionic Lagrangian
\begin{eqnarray}
\label{apsi}
L \, = \,  {1 \over 2} \, (\partial_{\mu}a)^2 \,  - \, V(a)\,
+ \, q_{eff} \partial^{\mu} a \epsilon_{\mu\nu}\bar{\psi}\gamma^{\mu}\psi  \,  + \,  i\bar{\psi} \gamma_{\mu} D^{\mu} \psi\, 
- \,  m\bar{\psi}\psi\, - \, {1 \over 4} F^2 
\end{eqnarray}
In the limit $m \,  \rightarrow \, 0,\, V(a)\,  \rightarrow \, 0$ there are the two continuous axial symmetries 
(\ref{ashiftc}) and 
\begin{equation}
\label{psishift}
\psi \, \rightarrow \, e^{i\gamma_5\theta} \psi
\end{equation}
But only the second is anomalous, and the corresponding speudo-Goldstone boson is 
the fermionic composite, and not $a$.  The composite pseudo-Goldstone can be  related to an
elementary scalar $\chi$ via standard bosonisation\cite{boson}
\begin{equation}
\label{bosonisation}
\bar{\psi}\gamma_{\mu}\psi\, = \, \epsilon _{\mu\nu}\partial^{\nu}{\chi \over \sqrt{\pi}}  ~~{\rm and} ~~ 
m\bar{\psi}\psi \, =\, \, m\kappa {\rm cos} 2\sqrt{\pi}\chi
\end{equation}
where $\kappa$ is a charge-related constant. 
The reason why the axial $U(1)$ symmetry cannot "see" $a$ as its pseudo-Goldstone boson  
has to do with the peculiarities of $1+1$ dimensions, and in particular  the fact that the axial 
and vector currents are related through
\begin{equation}
\label{axialvector}
\epsilon_{\mu\nu} \bar{\psi}\gamma^{\nu}\psi \, = \, \bar{\psi} \gamma_{\mu}\gamma^5\psi
\end{equation}

Generalization of the model (\ref{achi})  to four-dimensions is straightforward and the results are essentially unchanged.

\end{document}